# Safety of magnetic resonance imaging in patients with retained cardiac leads

Bach T. Nguyen[1], Bhumi Bhusal[1], Amir Ali Rahsepar[1], Kate Fawcett[1], Stella Lin[1], Daniel S Marks[2], Rod Passman[2], Donny Nieto[1], Richard Niemzcura[1], and Laleh Golestanirad[1,3]

[1] Department of Radiology, Feinberg School of Medicine, Northwestern University, Chicago, IL

[2] Department of Electrophysiology, Feinberg School of Medicine, Northwestern University, Chicago, IL

[3] Department of Biomedical Engineering, McCormick School of Engineering, Northwestern University, Evanston, IL

Corresponding author: Laleh Golestanirad, Department of Radiology, Feinberg School of Medicine, Northwestern University, 737 N Michigan Ave, Suite 1600, Chicago, IL 60611, USA. Email: Laleh.rad1@northwestern.edu

**Abstract**

*Purpose:* To evaluate safety of MRI in patients with fragmented retained leads (FRLs) through numerical simulation and phantom experiments.

*Methods:* Electromagnetic and thermal simulations were performed to determine the worst-case RF heating of 10 patient-derived FRL models during MRI at 1.5 T and 3 T and at imaging landmarks corresponding to head, chest and abdomen. RF heating measurements were performed in phantoms implanted with reconstructed FRL models that produced highest heating in numerical simulations. The potential for unintended tissue stimulation was assessed through a conservative estimation of the electric field induced in the tissue due to gradient-induced voltages developed along the length of FRLs.

*Results:* In simulations under conservative approach, RF exposure at $B_1^+ \leq 2\mu T$ generated $CEM_{43}$<40 at all imaging landmarks at both 1.5 T and 3 T indicating no thermal damage for acquisition times (TA) <10 minutes. In experiments, the maximum temperature rise when FRLs were positioned at the location of maximum electric field exposure was measured to be 2.4 °C at 3 T and 2.1 °C at 1.5 T. Electric fields induced in the tissue due to gradient-induced voltages remained below the threshold for cardiac tissue stimulation in all cases.

*Conclusions:* Simulation and experimental results indicate that patients with FRLs can be scanned safely at both 1.5 T and 3 T with majority of clinical pulse sequences.

## INTRODUCTION

With an aging population and the rising prevalence of cardiac disease, cardiovascular implantable electronic devices (CIEDs) are used more frequently. Today, there are more than 3 million Americans with CIEDs (1), and the number grows by 80,000 annually (2). It is estimated that 50-75% of these patients may need MRI for cardiac or non-cardiac indications (3), with many patients requiring repeated examination (4).

Several studies have assessed safety of MRI in patients with CIED leads connected to working devices. Some patients, however, require procedures which lead to "lead extraction", leaving fractions of the device *in situ* in which case the original safety studies of the intact device are not applicable. For example, patients undergoing heart transplant usually have a CIED in place, often with multiple leads in the vascular space (5). At the time of surgery the physician attempts to remove the device by cutting the leads and extracting the pulse generator, (leaving behind an exposed metal contact). Complete lead removal, however, is not always possible especially when there is adhesion of leads to the vessel wall or significant calcification (5). This leaves a sizeable cohort of patients with fragmented retained leads (FRLs) (6,7). MRI is needed in majority of these patients either for cardiac indications (e.g., rejection monitoring, perfusion assessment) or neurological and orthopedic exams. Today, however, there is no consensus on safety of MRI in patients with retained cardiac leads. Although MR-conditional CIEDs have been available since 2011, the conditional labeling of these devices applies only to the intact device and leads in their originally intended configuration. As such, studies assessing MRI safety of pacemaker or ICD generators mostly exclude patients with FRLs (8,9).

Due to lack of guidelines, the decision whether to perform MRI on a patient with a fragmented retained lead falls upon the physician's judgment, and is often based on her assessment of risk to benefit ratio. The goal of this work is to provide a comprehensive assessment of MRI hazards in patients with FRLs in order to provide actionable information to clinicians encountering such cases. We expanded our previous work which assessed SAR amplification around patient-derived FRL models in a single homogenous body model (10), by performing a thorough search for the worst-case heating scenario in body models with 30 unique combination of permittivity and conductivity covering a range of low to high values reported for biological tissues. Additionally, we included new FRL models with folded trajectories and double leads in close proximity as these topologies are shown to alter RF heating (11). To provide action points for clinicians, a conservative estimation of $CEM_{43}$ for different exposure times and $B_1^+$ values was provided along with thresholds of tissue thermal damage. We also performed experiments measuring RF heating around fragments of commercial CIED leads implanted in gel phantoms to gain additional confidence as to typical level of RF heating around FRLs *in vitro*. Finally, we assessed other sources of MRI hazard including the potential for unintended tissue stimulation.

**METHODS**

*Sources of MRI hazard in patients with FRLs*

The technical specification ISO-TS 10974 describes known sources of potential hazardous interaction between MRI fields and an active implantable medical device (AIMD). These include risk of gradient-induced and RF-induced device heating, potential harm due to gradient-induced vibration, gradient-induced extrinsic potentials, and gradient-induced device malfunction, as well as risks associated with $B_0$-inserted force and torque (12).

Gradient-induced device heating is the result of time varying imaging gradient dB/dt inducing eddy currents on AIMD conductive surfaces such as enclosures, battery components and circuit traces. For AIMDs with extended leads that do not contain larger conductive surfaces, there is no known mechanism for MRI-induced eddy current heating to occur in the lead (ISO-TS 10974:9). Therefore, this is not a source of concern in patients with FRLs.

Gradient-induced eddy currents also generate a time-varying magnetic moment that interacts with the static magnetic field ($B_0$) causing vibration of conductive surfaces (and subsequently the device) which could lead to device malfunction (ISO-TS 10974:10). For same reasons explained above, this is not a source of concern in patients with fragmented retained leads. Finally, the risks associated with lead traction and dislodgment due to $B_0$-inserted force and torque are negligible as CIED leads are composed of non-magnetic material. RF-induced heating and gradient-induced extrinsic voltages are therefore the only potential sources of MRI hazard in patients with FRLs.

*RF heating*

RF heating is due to the *antenna effect*, where the electric field of MRI transmit coil couples with conductive leads and amplifies the specific absorption rate (SAR) of radiofrequency energy in the tissue, usually occurring around the tip of an elongated implant (10,13-17). Temperature rises up to $\Delta T$ =30°C have been reported at tips of abandoned cardiac leads in phantom experiments at 1.5 T (18,19).

MRI RF heating of an elongated lead is known to be a resonance phenomenon, depending on lead's length and the distribution of MRI electric field's phase along lead's trajectory (13,20,21). We created clinically relevant FRL models based on medical images of 10 representative patients, including trajectories with folds or multiple leads in close proximity as they may alter RF heating (19,22-24).

Another important factor influencing RF heating is the electrical property of the medium surrounding the lead (25-28). To determine the worst-case scenario, we performed numerical simulations with each FRL model registered to homogenous body models with 30 unique combinations of permittivity and conductivity covering a range of low to high values reported for biological tissues ($\sigma \epsilon [0.1, 1]\ S/m, \varepsilon_r \epsilon [40, 80]$). The rationale for using a homogenous body model with varying electrical properties was based on a recent study that showed such model can predict the worst-case SAR generated in a heterogeneous body mode at a reduced computational cost (29). Finally, the position of the lead within the MRI RF coil affects its RF heating (30,31). Thus, RF heating was assessed for each FRL+ body model positioned inside the MRI body coil at landmarks corresponding to head, chest, and abdomen imaging.

Temperature rise in the tissue was conservatively estimated by solving the simplified Penne's bio-heat equation excluding cooling effects of the perfusion:

$$c\rho \frac{\partial T}{\partial t} - \nabla k \nabla T = \rho(SAR)$$

where $T$ is the temperature, $\rho$ is the density (1000 kg/m3), $c$ is the specific heat capacity of the tissue (4150 Jkg$^{-1}$°C$^{-1}$) and $k$ is the isotropic thermal conductivity (0.42 W/m$^{-1}$°C$^{-1}$) (32). For each FRL model, thermal simulations were performed in the human body model that generated the highest local SAR at the corresponding imaging landmark (abdomen, chest and head), with the input power of coils adjusted to produce $B_1^+$ in a range of 1 µT to 5 µT. The selected $B_1^+$ values cover the whole range of routine clinical protocols including high-SAR sequences.

To provide actionable information, we calculated the worst-case cumulative equivalent minutes at 43 °C (CEM$_{43}$) for a range of exposure times (1-10 minutes) and $B_1^+$ values (1µT-

5 µT) at 1.5 T and 3 T. CEM$_{43}$ is currently the accepted metric for thermal dose assessment that correlates well with thermal damage in variety of tissues (33,34). The calculation of CEM$_{43}$ requires knowledge of thermal history as:

$$CEM_{43} = \Delta t \times R^{(43-T)}$$

where $\Delta t$ indicates integration over the length of exposure, T is the average temperature during the exposure time, and R is a constant equal to 0.25 for T<43°C and 0.5 for T>43°C (33). To be conservative, CEM$_{43}$ was calculated using temperature rise ΔT from the FRL model that generated the maximum heating. Finally, we performed experiments with fragments of a commercial cardiac lead implanted in a gel phantom during RF exposure at 1.5 T and 3T to gain additional confidence as to level of RF heating observed *in vitro*.

*Gradient-induced voltage*

Extrinsic electric field potentials are gradient-induced voltages that develop between spatially separated electrodes within a single lead or between electrodes of a multi-lead system (ISO-TS 10974:13). If the FRL is in contact with the excitable tissue, these voltages could potentially cause unintended stimulation. We assessed the possibility of unintended tissue stimulation due to gradient-induced extrinsic potentials along FRLs based on a conservative estimation of electric potentials developed along the length of FRL models as described in ISO-TS 10974:13. These voltage values were then used to calculate a conservative estimation of electric field E induced in the tissue developed between the two ends of each FRL and the potential for unintended tissue stimulation was assessed in the context of cardiac stimulation thresholds (35,36).

*Patient-derived FRL models*

Chest CT and X-ray images of 100 patients with a history of implanted cardiac devices who had been admitted to Northwestern Memorial Hospital between 2006 and 2018 were obtained through a search of the Northwestern Universities' Enterprise Data Warehouse. Images were inspected by a radiologist (AR) for the presence of FRLs. From patients identified with FRL, 10 patients who had both chest X-ray and CT images that clearly delineated FRL trajectory and topology were included in the study. Patient characteristics are given in Table 1.

Lead trajectories were semi-automatically segmented from CT images using Amira (Amira 5.3.3, FEI Inc.), by applying a thresholding algorithm based on an intensity histogram analysis which extracted a preliminary mask of the hyper-dense lead from the CT image (Figure 1A). Lead centerlines were manually extracted and exported to a CAD tool (Rhino 3D, Robert McNeal and Associates, Seattle, WA) where 3D models of FRLs were constructed around them. CIED leads usually include individual coiled wires extending from the proximal to the distal end of the lead. The topology of these coiled wires changes during the extraction process as the physician applies force to extract the lead due to surrounding adhesions. We used X-ray images to reconstruct the details of the FRL structure, including the number of micro-loops and the variation in their pitch (Figure 1B-C). To correctly position the FRLs in the human body model, we created a triangulated surface of each patient's ribcage from CT images, which was then used to align and co-register the FRL of that patient to the ANSYS multi-compartment human body model (37) (Figure 1D). Once the FRL was in its correct position, we merged different body tissues to create a homogenous body model, which was then assigned to a range of different electrical conductivity ($\sigma$ = 0.1, 0.2, 0.4, 0.6, 0.8 and 1 *S/m*) and permittivity ($\varepsilon_r$ =40, 50, 60, 70 and 80) to cover all relevant biological values (38). This is reasonable, as a recent study has showed a homogeneous body model with conductivities varied in a range of low to high values (0.01 *S/m* $\rightarrow$ 1 *S/m*) can predict the full range of SAR variations predicted by a heterogeneous body model (29).

*MRI RF coils*

Numerical models of two high-pass birdcage body coils (620 mm length, 607 mm diameter) were implemented in ANSYS Electronic Desktop and tuned to their respective Larmor frequencies - 64 MHz (1.5 T) and 127 MHz (3 T). Each RF coil was composed of 16 rungs connected at each end to two end rings and shielded by an open cylinder (1220 mm length, 660 mm diameter). The coils and shields were made of copper and their dimensions were chosen based on a typical clinical body coil. A quadrature excitation was implemented by feeding the coils at two ports on the bottom end-ring that were 90° apart in position and phase. Coils were tuned by lumped capacitors distributed at the end-ring gaps (86.5 pF for 64 MHz and 17 pF for 127 MHz) and matched to 50 Ω using a single capacitor at each port (100 pF at 64 MHz and 45.6 pF at 127 MHz) in series with an ideal voltage source with a 50 Ω internal resistance.

*SAR and thermal simulations*

A total of 1800 numerical simulations were performed, with 10 realistic FRL models incorporated in body models with 30 unique combination of electrical conductivity and permittivity, positioned in 2 MRI body coils (1.5 T and 3 T) at 3 different imaging landmarks (abdomen, chest, head). The maximum value of 1g-averaged SAR (referred to as MaxSAR1g) was calculated inside a conformal cylindrical volume (diameter = 2 or 3 *cm* depending on the shape of the FRL) that surrounded the FRL as illustrated in Figure 2.

To ensure good numerical convergence, the initial mesh was set such that the maximum element size was <2 mm on the FRL, <2 mm in the surrounding conformal region in which MaxSAR1g was calculated, and <50 mm in the body. ANSYS HFSS was set to follow an adaptive mesh scheme with successive refinement of the initial mesh between iterative passes until the difference in magnitude of the S-parameters fell below a set threshold of 0.02. Total time taken for each simulation was from thirty minutes to two and half hours (depending on the length and the shape of FRL) on a DELL server with 1.5 TB memory and 2_Xenon(R) Gold 6140 CPUs each having 18 processing cores. Table 2 gives mesh statistics for a typical simulation.

Thermal simulations were performed in the human body model that produced the highest local SAR at each imaging landmark (abdomen, chest and head), with input power of coils adjusted to produce a spatial mean of $B_1^+$ amplitude (i.e., complex magnitude of $B_1^+$ averaged over an axial plane passing through iso-center of the coil, see Figure 2) in a range of 1 µT to 5 µT.

*RF heating measurements*

To gain additional confidence as to typical values of RF heating occurring in the tissue surrounding FRLs, we performed phantom experiments with four FRL models reconstructed from commercially available pacemaker leads (Medtronic 5076), with trajectories mimicking those that generated highest temperature rise in simulations. Specifically, we reconstructed FRL #6 which generated highest temperature rise at 1.5 T for all three landmarks, and FRL models #4, #3 and #9 which generated highest temperature rise at 3 T at abdomen, chest and head landmarks, respectively (see Figure 3).

Experiments were performed in a Siemens Aera 1.5 T scanner and a Siemens Prisma 3 T scanner (Siemens Healthineers, Erlangen, Germany). FRL models were positioned inside a human-shaped container filled with a gel (12 cm thick) prepared by mixing 10 g/L of Polyacrylic Acid (PAA, Aldrich Chemical) and 1.32 g/L Sodium Chloride with distilled water. The anthropomorphic phantom was designed using segmentation of patient images as described in our previous work (27). Bulk conductivity and relative permittivity of the gel was measured to be σ = 0.46 S/m and $ε_r$ = 87 using a dielectric probe kit (85070E, Agilent Technologies, Santa Clara, CA) and a network analyzer.

RF heating was measured for two scenarios. First, FRL models were positioned in a location analogous to the middle of the chest similar to what we observed in patient images, namely position P1 as illustrated in Figure 6A. Second, and to be more conservative, we also measured temperature rise for FRL models in a high field exposure location, namely position P2 in Figure 6A which was near the phantom wall. To find the location of maximum field exposure, we performed experiments with a simple insulated copper wire (see Supporting Information Figure S6) which was orientated parallel to the phantom's long walls at three different depths from the gel's surface. The temperature profile showed a right-left asymmetry in local E-field distribution, which was in agreement with the earlier findings (39). The highest temperature rise was occurred when the wire was located along the phantom's left wall and 2 cm from the gel's surface. This location was chosen as P2 to position the FRL at the maximum E field exposure. Two fluoroptic temperature probes (OSENSA, BC, Canada) were secured at both tips for all 4 FRL models. Temperature rise ∆T in the gel was recorded during 10 minutes of RF exposure using a high-SAR T1-turbo spin echo sequence (TE = 7.3 msec, TR=814 msec, flip angle = 150° for 1.5 T scans and TE = 7.5 msec, TR = 1450 msec, flip angle = 150° for 3 T scans). An additional temperature probe was also positioned inside the phantom far from the implant position, to observe the background heating (Figure 6). The sequence parameters were adjusted to generate the RMS $B_1^+$ of 2µT in the phantom. Imaging was performed using scanner's built-in body coil. For each FRL model, the phantom was positioned alternately such that the abdomen/chest/head of the phantom was at the coil's iso-center. For all experiments, precise positioning of the FRLs inside the gel was ensured by using the support pillars with adjustable height which were positioned on a gird fitted to the bottom surface of the anthropomorphic phantom. Figure 6A shows details of the experimental setup.

*Estimation of gradient-Induced extrinsic voltages*

To obtain a conservative estimation of electric field induced in the tissue, we first calculated the injected $V_{emf}$ along each FRL using the Tier 1 approach described in ISO-TS10974:13 (12). This approach is based on simulated values of electric fields induced on the surface of a cylinder of 20 cm radius by a dB/dt=100 T/s, and gives the most conservative estimation of extrinsic voltages for leads with length < 63 cm. We then estimated the electric field E induced in the tissue between the two ends of each FRL by dividing the $V_{emf}$ by the distance between the end tips (see Supporting Information Figure S7). These values were compared against cardiac stimulation (CS) thresholds provided in the literature (40-42).

# RESULTS

*SAR distribution and maximum temperature rise in the tissue*

Figure 3 gives MaxSAR1g values for each FRL model during RF exposure at 64 MHz (1.5 T) and 127 MH (3 T) and for the body model positioned at head, chest, and abdomen landmarks. At each imaging landmark, the input power of the coil was adjusted such that the spatial mean of $B_1^+$ amplitude was 2µT at coil's iso-center. Boxcars represent the variation of MaxSAR1g as a function of electrical properties ($\sigma$ and $\varepsilon_r$) of the body model. As it can be observed, the specific combination of tissue permittivity and conductivity that produced maximum SAR was different for each FRL model, emphasizing the interdependency of different factors affecting RF heating (24). Plots of variation of MaxSAR1g as a function of conductivity and permittivity are given in the Supporting Information Figures S1-S5. In general, lower values of conductivity generated higher SAR in most models, with the trend being substantially more pronounced at 3 T.

MaxSAR1g (mean ± standard deviation) was 3.50±4.23 W/kg for the head imaging landmark, 11.34±14.20 W/kg for the chest imaging landmark and 3.20±6.56 W/kg for the abdomen imaging landmark for RF exposure at 1.5 T (averaged over all body compositions and all lead models, n=300). MaxSAR1g was significantly higher at chest imaging landmark compared to both head landmark and abdomen landmark (one-tail t-test, p-value<1E-04). There was no significant difference between MaxSAR1g at head and abdomen landmarks (two-tail t-test, p-value=0.5).

At 3T, MaxSAR1g was 11.79±14.52 W/kg for the head imaging landmark, 32.09±34.65 W/kg for the chest imaging landmark and 6.83±17.92 W/kg for the abdomen imaging landmark (averaged over all body compositions and all lead models, n=300). MaxSAR1g was significantly higher at chest imaging landmark compared to both head landmark and abdomen landmark (one-tail t-test, p-value<1E-04). MaxSAR1g at head landmark was slightly higher MaxSAR1g at abdomen landmark (one-tail t-test, p-value=1E-04). At all imaging landmarks, MaxSAR1g was significantly higher at 3 T compared to 1.5 T (one-tail paired t-test, p-values<1E-04).

Figure 4 shows histograms of MaxSAR1g distribution at each field strength and for each imaging landmark. An exponential probability density function (mean of $\mu$) was fitted to each distribution using the MATLAB and Statistics Toolbox Release 2019a (The MathWorks, Inc., Natick, Massachusettes). For each case, the model that created the most extreme data point on the SAR distribution was used for subsequent thermal analyses. From Figure 4, it can be observed that approach was conservative as >99% of cases generated SAR values below this limit.

*Thermal dose*

The temperature rise in the tissue after a 10-minute of continuous RF exposure at 1.5 T and 3 T is given in Supporting Information Tables S1 and S2, respectively. To be conservative, thermal simulations were performed using the body model with electrical properties that generated the highest MaxSAR1g.

At each imaging landmark and for each $B_1^+$ value, we calculated $CEM_{43}$ corresponding to the FRL model that generated the highest RF heating (see Figure 5). A review of published data on tissue damage due to thermal exposure can be found in (34). For muscle tissue, $CEM_{43}$>80 min has been reported to cause chronic damage, whereas 41<$CEM_{43}$<80 min was associated with acute but minor damages. From Figure 5 it can be observed that for $B_1^+ \leq 2\mu T$, RF exposure at both 1.5 T and 3 T generated $CEM_{43}$<40 at all imaging landmarks, indicating no thermal damage for acquisition times (TA) <10 minutes.

It should be noted, however, that calculated $CEM_{43}$ values for continuous RF exposure are highly conservative as MRI sequences typically have duty cycles much below 100%. Table 3 give examples of typical clinical sequences for neuroimaging, cardiac imaging, and body imaging with their corresponding $B_1^+$ and acquisition times. The sequence-specific $CEM_{43}$ calculated for the FRL model that generated the maximum SAR is also given. As it can be observed $CEM_{43}$ remained well below 40 for all sequences, indicating no thermal damage.

*Experimental measurements*

Figure 6B gives the maximum $\Delta T$ recorded along the length of each FRL model at each landmark and field strength. Imaging at chest landmark produced highest heating at both 1.5 T and 3 T, consistent with simulation predictions. In addition, higher heating was observed at 3 T compared to 1.5 T for most of the cases, which is also consistent with the simulation results. The maximum $\Delta T$ at high field exposure position (P2) was 2.4 °C for FRL #3 during MRI at 3 T and 2.1 °C for FRL #6 during MRI at 1.5 T. The corresponding values of maximum heating for realistic position (P1) were 0.7 °C for 3 T and 0.6 °C for 1.5 T.

*Estimated extrinsic voltages*

Table 4 gives the values of gradient-induced extrinsic potentials calculated from Supporting Information Table S3 (based on table A.3 in ISO-TS10794) and a conservative estimation of induced E field in the tissue between the two ends of each FRL. The strength-duration parameters estimated by Reilly (40) form the basis for the CS limit defined in the IEC 60601-2-33 safety guidelines (35). With a rheobase of 6.2 V/m and a time constant of 3 ms, Reilly calculated the CS limit as:

$$E < \frac{6.2 \ V/m}{1 - \exp{(-\frac{\tau_{eff}}{3 \ ms})}}$$

where $\tau_{eff}$ is the effective length of the stimulation, which in this case is the rise time of the gradient pulse. Most clinical scanners have a gradient rise time of 0.1 ms, leading to E field threshold of 189 V/m. From Table 4, all induced E values remained below this threshold.

**Discussion and Conclusion**

There is a steady growth in the use of cardiac implantable electronic devices (CIED) in the United States and globally. The trend is likely to continue with new indications for use and technological advancements in device manufacturing (43,44). Despite the increasing need, MRI is still largely inaccessible to patients with CIEDs because of safety hazards underscored by several injuries reported worldwide (45-48). Although new generations of pacemakers and defibrillators have reduced safety risks associated with MRI static and gradient fields affecting device function, tissue heating from the radiofrequency excitation fields remain a major issue.

This so-called "antenna effect" happens when the electric field of the MRI transmitter couples with implanted leads of the CIED, causing the specific absorption rate (SAR) of the RF energy to significantly amplify at the implant's tip. In vitro and in vivo studies report temperature rises up to 20º C at the lead tip highlighting the significance of the issue (49,50).

Efforts to develop MRI-compatible CIEDs are recent, with newly approved devices allowing conditional MRI of patients with intact cardiac leads at both 1.5 T and 3 T scanners. There is, however, a sizeable cohort of patients with abandoned or retained leads for whom MRI under current labeling is an absolute contraindication, mostly because very little is known about the phenomenology of MRI-induced heating in the tissue in the presence of partially extracted leads. The major challenge in quantifying implant-induced heating in this case is that the problem has a large parameter space with many variables that interact with each other. This includes the frequency and geometry of MRI RF coil, the length, trajectory, and structure of the abandoned/retained lead, the imaging landmark, and the patient's anatomy. Such complexity precludes the application of a systematic experimental approach to infer the worst-case heating scenario. Numerical simulations on the other hand, provide an exquisite methodology for exploring thousands of variable combinations in a holistic manner, allowing analysis of parameter extremes outside the bounds of normal clinical practice. This work provides the first comprehensive simulation study of MRI hazards in patients with fragmented retained leads (FRLs) with a focus on RF heating and unintended tissue stimulation. We performed a total of 1800 simulations with 10 patient-derived FRL models registered to body models with a range of low to high relative permittivities ($\varepsilon_r$ =40-80) and conductivities ($\sigma = 0.1 - 1\ S/m$), positioned in MRI RF coils tuned to 64 MHz and 127 MHz, and at different imaging landmarks corresponding to head, chest, and abdomen imaging. In general, body models with lower conductivity generated higher SAR around FRL models. The trend was specifically pronounced at 3 T, where the maximum of SAR1g was almost always generated in the body model with $\sigma = 0.1\ S/m$ (with the exception of FRL #2 and FRL #7 at chest landmark, where body model with $\sigma = 0.2\ S/m$ and $\sigma = 0.6\ S/m$ generated the maximum SAR, respectively).

In terms of imaging landmark, we found an agreement in both simulation and experiment results that chest landmark produced significantly higher SAR compared to head and abdomen landmarks at both 1.5 T and 3 T. Specifically, at 1.5 T MaxSAR1g was ~4-folds higher during RF exposure at chest landmark compared to head and abdomen landmarks. Similarly, at 3 T, MaxSAR1g was ~3-folds and 5-folds higher at chest landmark compared to head and abdomen landmarks, respectively. This was predictable, considering that FRLs were almost exclusively located in the subclavian vein which would position them at the location of maximum RF field for the coil iso-center at chest. The experimental measurements predicted similar heating pattern in terms of variation in imaging landmarks and RF frequency. RF heating at both 1.5 T and 3 T MRI remained below 3 °C for all the FRL models used in the measurement.

Although SAR has been used as a surrogate to assess risks of RF heating of implants, temperature rise in the tissue is the ultimate indicator of tissue damage. Here we used a highly conservative approach to calculate the temperature rise $\Delta T$ in the tissue around each FRL model by (a) using the body model that generated the maximum SAR around each FRL model, (b) eliminating cooling effects of perfusion from Penn's bioheat equation, and (c) calculating the $\Delta T$ for a continuous RF exposure, despite the fact that almost all MRI sequences have duty cycles well below 100%. Another level of conservatism was applied in determination of CEM$_{43}$, which was calculated using $\Delta T$ from the FRL model which produced the maximum RF heating. RF exposure with $B_1^+ \leq 2\mu T$ generated $CEM_{43}$<40 at all imaging landmarks indicating no thermal damage for acquisition times (TA) <10 minutes. It should be noted, however, that many clinical sequences with higher $B_1^+$ are much shorter than 10 minutes, and thus, generate negligible $CEM_{43}$. Finally, a highly conservative assessment of electric field induced in the tissue due to gradient-induced voltages induced along the length of the FRLs suggested that the risk of unintended tissue stimulation was negligible.

Although MRI is currently contraindicated in patients with fragmented retained leads, recent studies that retrospectively analyzed available radiographic images and clinical records found no adverse effect associated with application of MRI in patients with FRLs (51). Our work provides theoretical and experimental evidence that MRI, under certain conditions, namely, for RMS $B_1^+ \leq 2\mu T$, may be performed safely in patients with FRLs and as such can serve as a guideline for future applications. These findings suggest that in patients with fragmented retained lead who their clinical management would be changed based on the diagnostic power of MRI, and there is no alternative to MRI such as CT imaging, MRI with caution may be performed without significant adverse effect, however, for routine use of MRI in these patients, larger cohort of patients must be studied.

Finally, we would like to emphasize the difference between "abandoned" and "retained" leads. Although the words are sometimes used interchangeably, they refer to substantially different scenarios in terms of MRI RF heating. Abandoned leads are leads that are simply disconnected from the CIED generator and left in the body. As such, their internal geometry and trajectory remains intact which means conductive wire remain within the insulation with only the tip exposed to the tissue. Retained leads on the other hand, are fragments of bare conductive wires (the insulation comes off during the extraction) that are left in the body after an attempt is made to extract the lead. As RF heating of insulated leads is in general higher than RF heating of bare leads the results of this study should not be generalized to assume safety of MRI in patients with abandoned leads or any FRL lead that has insulation retained with it.

**Acknowledgement**

This work was supported by NIH grant R03EB025344.

The authors would like to thank Dr. Sunder Rajan from the FDA Office of Science and Engineering Laboratories, Center for Devices and Radiological Health, for helpful discussions and guidance.

Supporting Information Table S1. Temperature rise $\Delta T$ [°C] in the tissue surrounding the FRL after 10-minute continuous RF exposure at 64 MHz (1.5 T) for the coil iso-center positioned at different imaging landmarks and the input power adjusted to generate different B1+ values on an axial plane passing through center of the coil.

Supporting Information Table S2. Temperature rise $\Delta T$ [°C] in the tissue surrounding the FRL after 10-minute continuous RF exposure at 127 MHz (3 T) for the coil iso-center positioned at different imaging landmarks and the input power adjusted to generate different B1+ values on an axial plane passing through center of the coil.

Supporting Information Table S3. Lead length factor L.

Supporting Information Figure S1. MaxSAR1g generated around retained lead models FRL1 and FRL2 as a function of body model's conductivity (horizontal axis) and permittivity (different-colored graphs). The field strength and imaging landmark is noted on top of each plot.

Supporting Information Figure S2. MaxSAR1g generated around retained lead models FRL3 and FRL4 as a function of body model's conductivity (horizontal axis) and permittivity (different-colored graphs). The field strength and imaging landmark is noted on top of each plot.

Supporting Information Figure S3. MaxSAR1g generated around retained lead models FRL5 and FRL6 as a function of body model's conductivity (horizontal axis) and permittivity (different-colored graphs). The field strength and imaging landmark is noted on top of each plot.

Supporting Information Figure S4. MaxSAR1g generated around retained lead models FRL7 and FRL8 as a function of body model's conductivity (horizontal axis) and permittivity (different-colored graphs). The field strength and imaging landmark is noted on top of each plot.

Supporting Information Figure S5. MaxSAR1g generated around retained lead models FRL9 and FRL10 as a function of body model's conductivity (horizontal axis) and permittivity (different-colored graphs). The field strength and imaging landmark is noted on top of each plot.

Supporting Information Figure S6. Measured temperature rise along length of 20 cm wire at 1.5 T. The wire was located at the left and the right of the phantom, and the depth of the wire was 2 cm, 5 cm and 8 cm from the top of the gel (maximum thickness = 12 cm).

Supporting Information Figure S7. (1): Two ends of the FRL. (2) Tangential component of gradient-induced electric field along the FRL. A conservative estimation of the induced voltage $V_{emf}$ along the FRL was calculated by multiplying the maximum value of gradient-induced E field (based on simulations given in Annex B of ISO-TS 10974) by the FRL length. (3) Gradient field (4) A conservative estimation of E field in the tissue was calculated by dividing $V_{emf}$ by the distance between two ends of the lead.

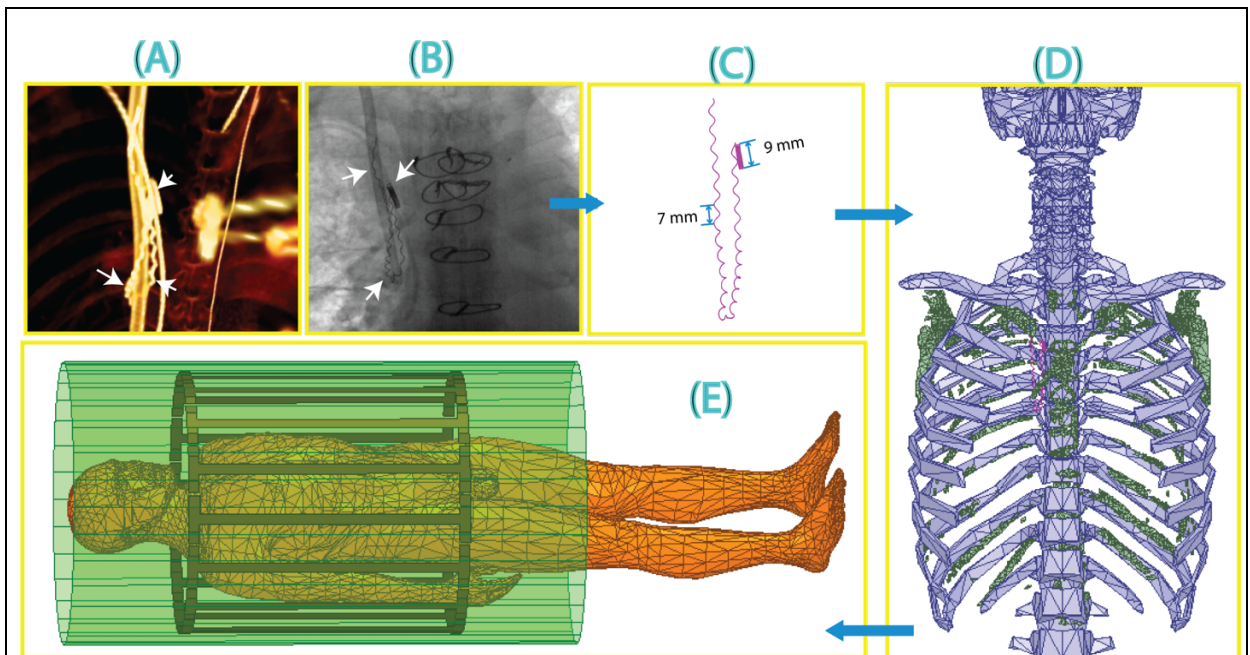

Figure 1. Steps of image segmentation and FRL model construction. (A-C) CT images were used to extract the 3D trajectory of FRL whereas x-ray images were used to reconstruct the FRL's structure (e.g., number and pitch of loops). White arrows show the FRL on each image. (D) A triangulated surface of patient's ribcage (green) was created and aligned with ANSYS multicompartment body model (blue) to position the patient-derived FRL model inside ANSYS human body model. (E) Body model positioned inside MRI coil.

Table 1. Characteristics of patients with CIED fragmented retain leads

| Patient ID# | Age | Sex | FRL Location | Device Type and Manufacturer | Lead Model(s) | Date of Implantation (m/y) | Date of Extraction (m/y) | Apparent/True Length of FRL [cm] |
|---|---|---|---|---|---|---|---|---|
| FRL1 | 24 | M | -Left subclavian vein<br>-SVC | Medtronic Dual chamber ICD | RA: Capturefix novis 5076<br><br>RV: Sprint Fidelis 6931 | 9/2007 | 1/2013 | 13.4 & 10.5<br>67.9 & 10.5 |
| FRL2 | 56 | M | -SVC/RA junction<br>-Left subclavian vein | Medtronic Biventricular ICD | RA: Capturefix novis – 5076<br>Spring quatro 6944 – RV<br>4193 Attain OTW - LV | 5/2002<br><br>9/2005 – LV | 04/2007 | 14.2<br>90.4 |
| FRL3 | 52 | F | -SVC<br>-Left subclavian vein | Boston Scientific Dual chamber ICD<br>Abbott St Jude RA lead | 1688 TC - RA<br>Reliance G 0185 - RV | 1/2007 | 08/2011 | 8.9 & 4.4<br>13.3 & 6.5 |
| FRL4 | 64 | F | -SVC/RA junction | Medtronic Biventricular ICD | Capture fix - 5076 – RA<br>Attain OTW– LV<br>Sprint Fidelis - RV | 4/2007 | 7/2008 | 18.0<br>41.2 |
| FRL5 | 66 | F | -SVC/RA<br>-Left subclavian vein | Medtronic Single chamber ICD | Sprint Quatro 6947-65 - RV | 4/2012 | 9/2012 | 17.2<br>107.3 |
| FRL6 | 57 | M | -SVC<br>-Left subclavian vein | Medtronic Biventricular ICD | Capturefix Novis 5076 – RA<br>Sprint Quatro 6947 – RV<br><br>Attain Starfix 4195 - LV | 4/2003<br><br>9/2008 – LV | 12/2009 | 15.5<br>54.3 |
| FRL7 | 50 | M | -SVC/RA junction | Medtronic Biventricular ICD | Capturefix novus 5076 – RA<br>Attain OTW 4194 – LV<br><br>Sprint Quatro 6947 - RV | 1/2006<br><br>3/2011 – LV<br><br>11/2011 - RV | 6/29/2015 | 2.0<br>26.4 |
| FRL8 | 57 | M | -SVC/RA | Medtronic Biventricular ICD | 5071 – LV<br><br>CaptureFix Novis 5076 – RA<br><br>Sprint Quatro Secure 6947 - RV | 8/2012 | 05/2013 | 4.5<br>46.9 |
| FRL9 | 64 | M | -SVC<br>-Left subclavian vein | Boston Scientific Dual chamber ICD | Boston Scientific 4052 – RA<br>Guidant Endotak Reliance 0185 – RV | 6/2003 | 11/2005 | 12.7 & 12.7<br>21.8 $ 18.1 |
| FRL10 | 29 | M | -SVC<br>-Left subclavian vein | Medtronic Biventricular ICD | CaptureFix Novis 5076 – RA<br>Sprint Fidelis 6949 – RV<br><br>Attain Ability 4196 – LV | 2/2005<br><br>10/2009 – LV | 08/2015 | 8.5<br>96.0 |

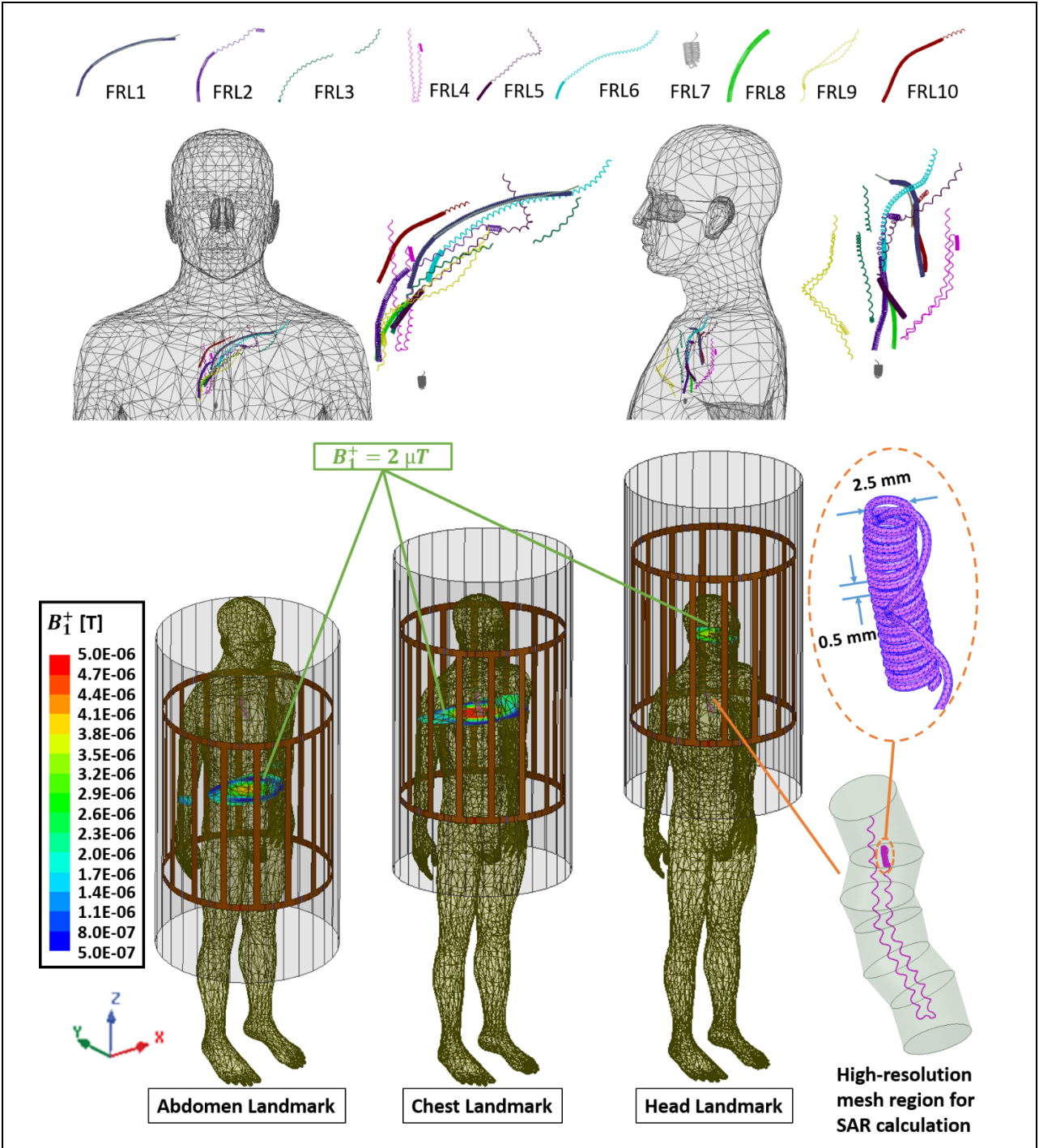

Figure 2. Top row: Trajectories and detailed structures of FRL models extracted from 10 patients (namely FRL1-FRL10). FRL 4&7 represent folded trajectories. FRL 9 was a case of a patient with two closely situated retained leads. Middle row: Front and side views of reconstructed FRLs and their relative locations in the human body model. Bottom row: Position of body model inside MRI coil at different imaging landmarks (namely abdomen, chest and head). The high-resolution mesh area around the lead in which 1g-SAR was calculated is also shown.

Table 2. Details of mesh statistics for the simulation of FRL4 at 3 T for the abdomen imaging landmark

| Parts | Num Tets | Min edge length [mm] | Max edge length [mm] | RMS edge length [mm] | Min tet. vol. [mm$^3$] | Max tet. vol. [mm$^3$] | Mean tet. vol. [mm$^3$] | Std devn vol. [mm$^3$] |
|---|---|---|---|---|---|---|---|---|
| Lead | 22597 | 0.0538 | 0.7218 | 0.4354 | 7.9218E-08 | 9.5171E-03 | 1.8853E-03 | 8.6679E-04 |
| SAR region | 506346 | 0.0566 | 2.3516 | 1.5475 | 7.7631E-08 | 7.6573E-01 | 1.7217E-01 | 1.0858E-01 |
| Body | 89169 | 1.0457 | 56.6824 | 24.927 | 8.3179E-03 | 9.7123E+03 | 8.7479E+02 | 1.2956E+03 |

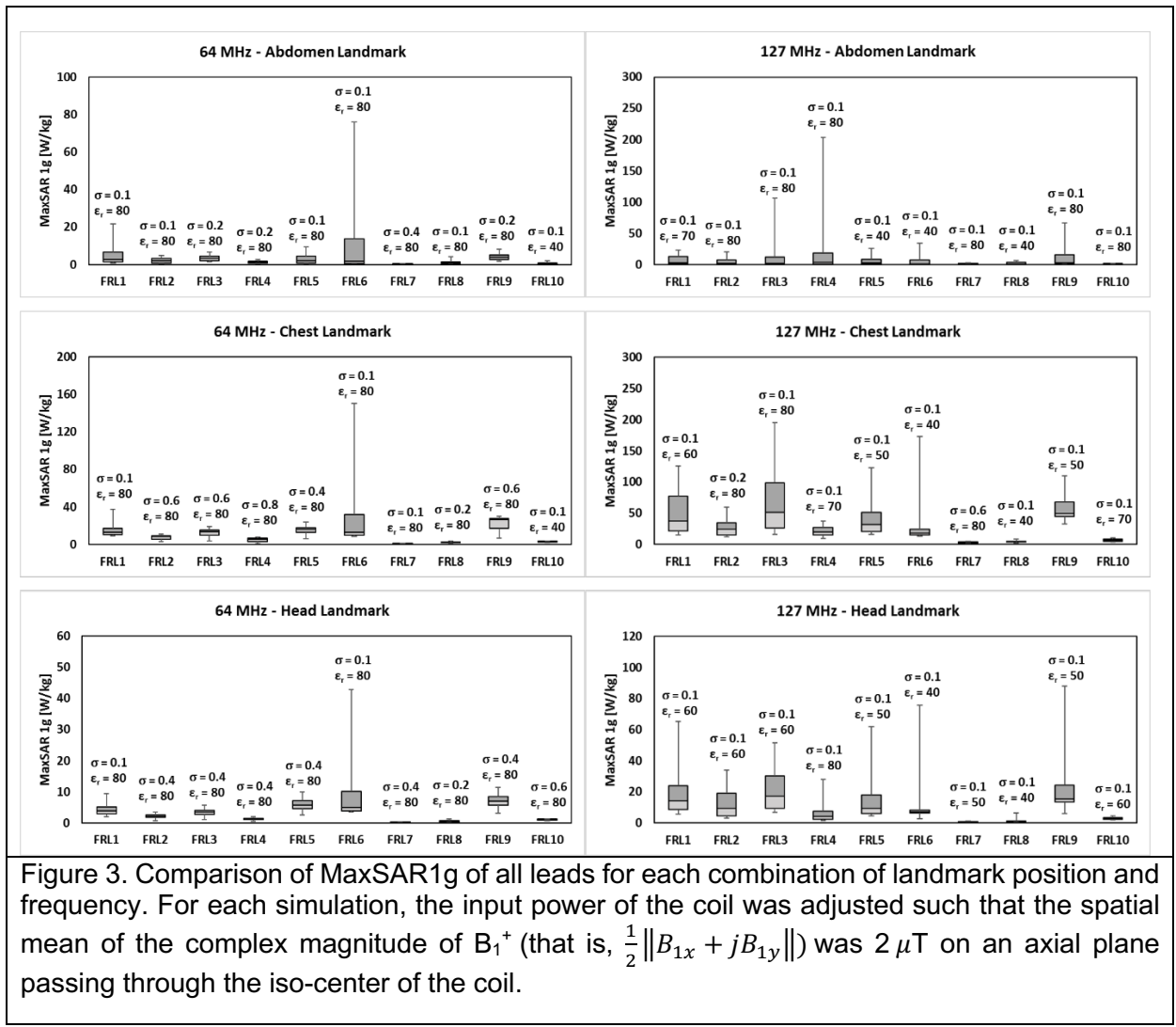

Figure 3. Comparison of MaxSAR1g of all leads for each combination of landmark position and frequency. For each simulation, the input power of the coil was adjusted such that the spatial mean of the complex magnitude of $B_1^+$ (that is, $\frac{1}{2}\|B_{1x} + jB_{1y}\|$) was $2\,\mu T$ on an axial plane passing through the iso-center of the coil.

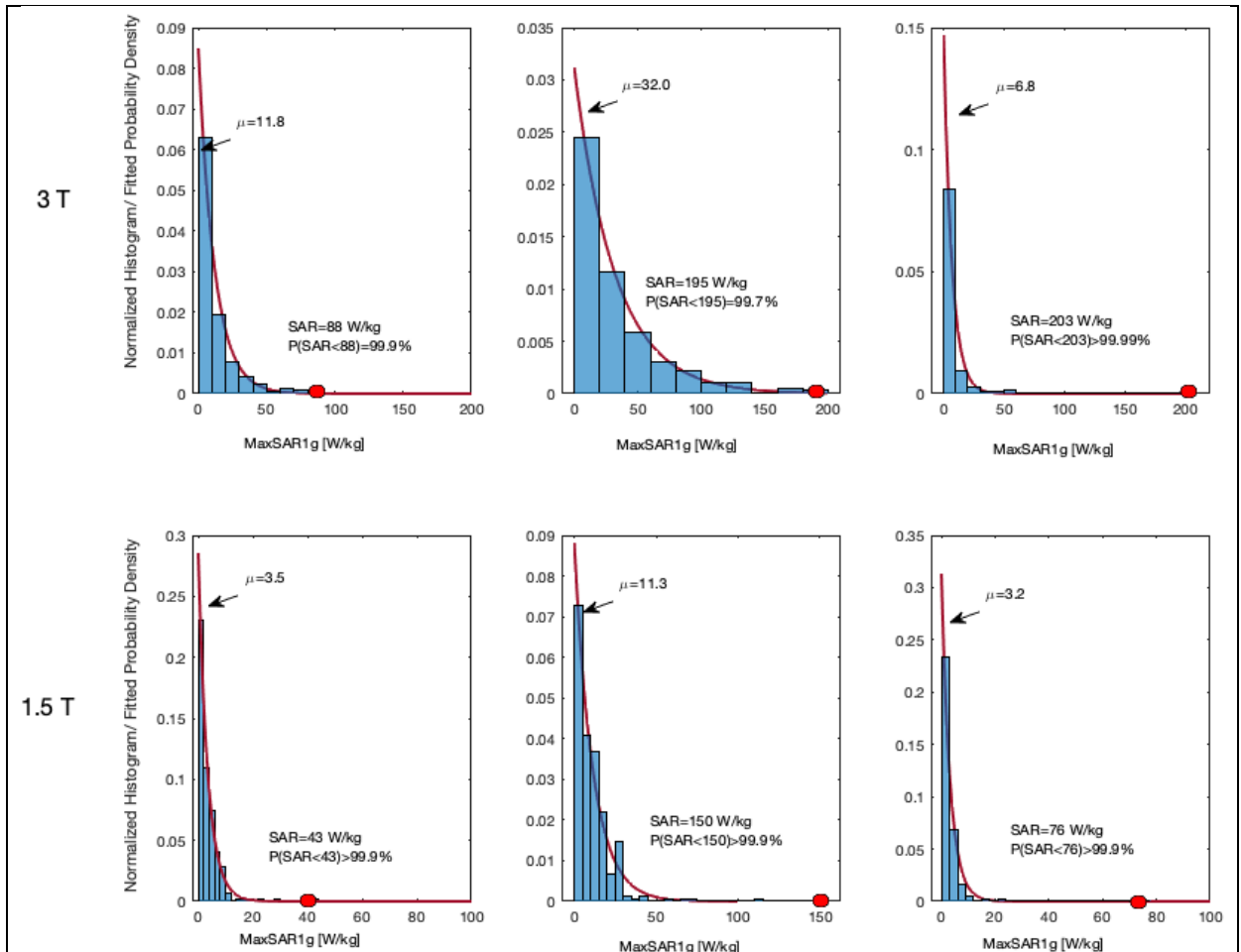

Figure 4. Normalized SAR histograms pooling N=300 body models and FRL leads for each field strength and at each imaging landmark. An exponential probability density function (mean=$\mu$) was fitted to the SAR distribution for each case. Red markers show the location of extreme cases associated with the body+FRL model that created the maximum SAR. The value of the maximum SAR and the probability of an individual SAR being less than this value are also given.

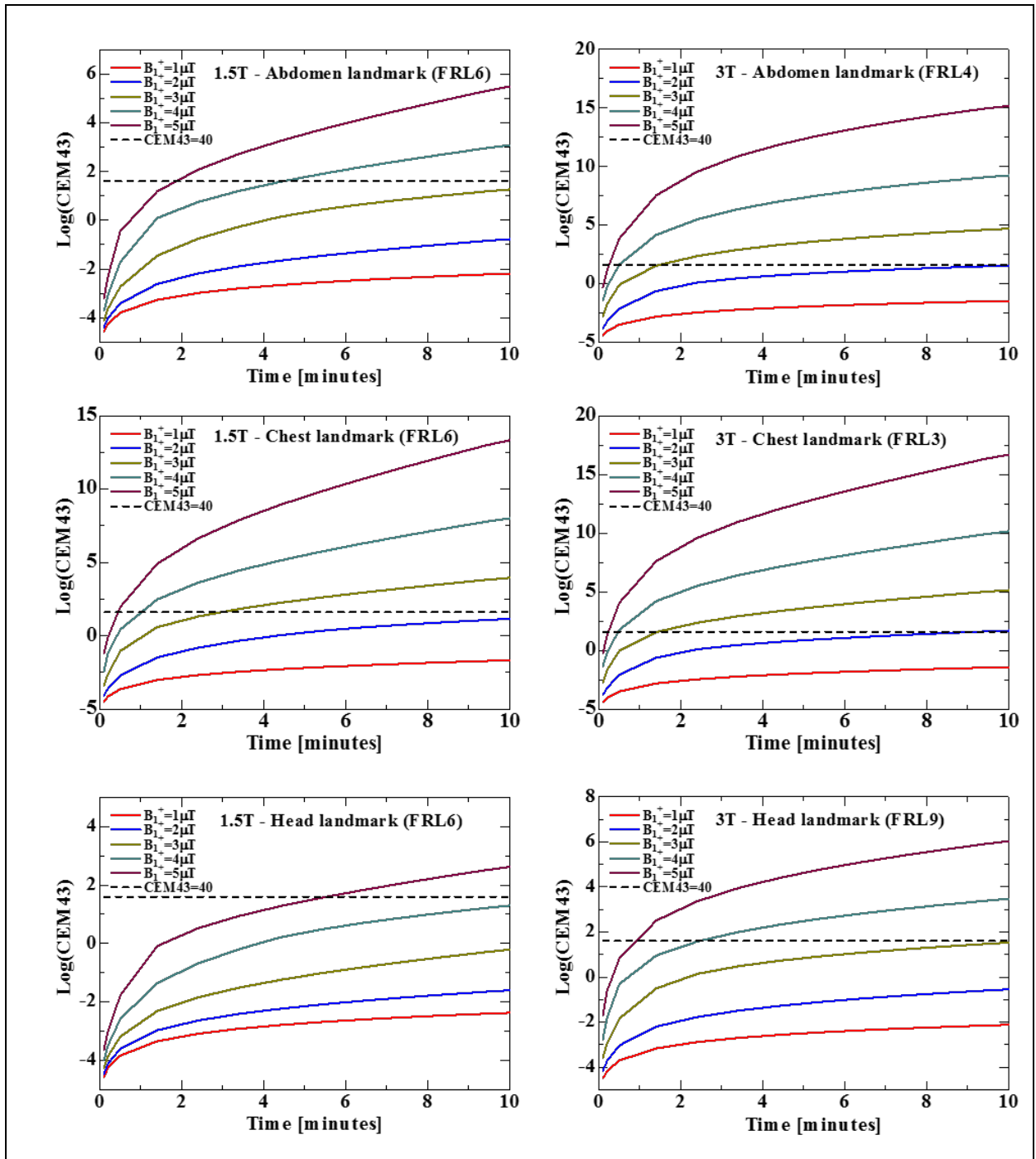

Figure 5. CEM$_{43}$ values calculated at different B$_1^+$ levels as a function of acquisition time. CEM$_{43}$ values are calculated for the FRL model that generated the maximum temperature rise at each imaging landmark and RF frequency. Reported B$_1^+$ values are spatial means of the complex magnitude of the B$_1^+$ calculated on an axial plane passing through coil's iso-center.

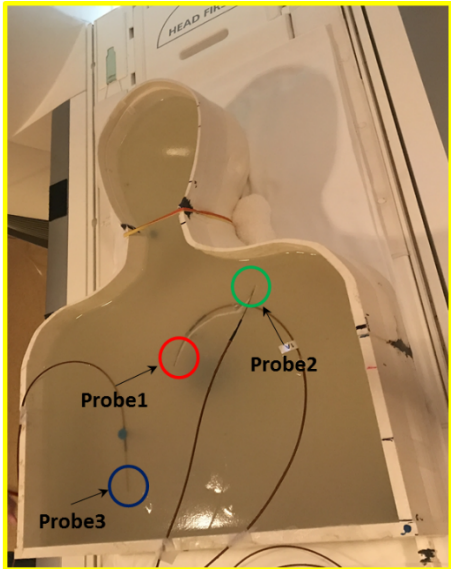
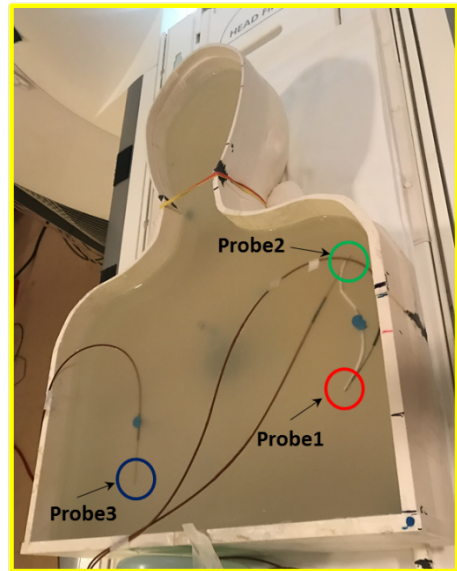

FRL3 is located at original position (P1)

FRL3 is located at high field exposure position (P2)

(A)

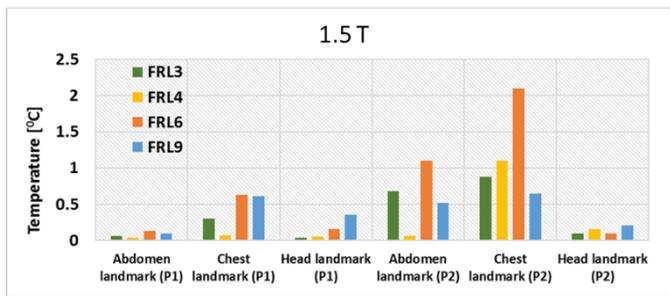
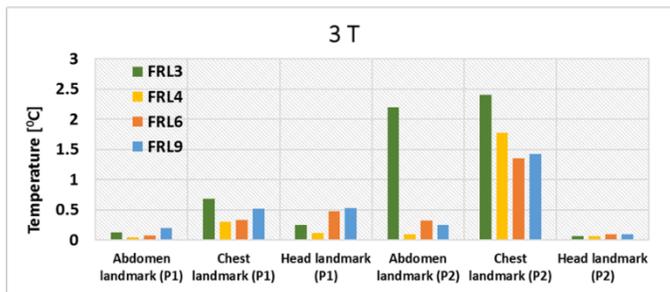
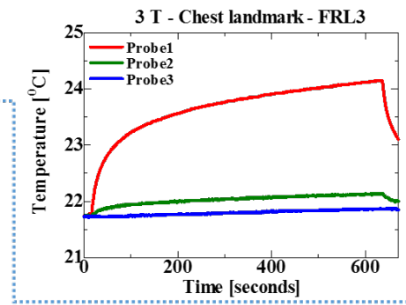
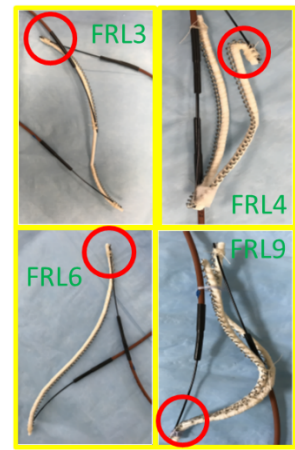

(B)

Figure 6. (A) The experimental setup for FRL3 and the positions of temperature probes, as well as their temperature responses at 3T at chest landmark at P2 case. (B) Maximum temperature rise along length of FRL models at 1.5T and 3T, as well as the locations where the maximum temperatures were measured.

Table 3. Sequence Parameters and scanner-reported RMS $B_1^+$ values for different imaging protocols at 1.5T scanner.

| 1.5 T Imaging | | | | | | |
|---|---|---|---|---|---|---|
| Head Landmark | | | | | | |
| Protocol | TE (ms) | TR (ms) | TA (min) | FA (°) | $B_1^+$ (µT) | $CEM_{43}$ |
| SE T1 SAG | 10 | 450 | 2:09 | 90 | 3.54 | 3.39E-02 |
| AX DIFFUSION | 89 | 7900 | 2:16 | - | 1.3 | 9.37E-04 |
| AX FLAIR | 86 | 9000 | 3:02 | 150 | 2 | 3.09E-03 |
| AX GRE T2 | 25 | 830 | 1:58 | 20 | 0.3 | 4.93E-04 |
| AX SWI | 40 | 49 | 3:38 | 15 | 0.3 | 8.32E-04 |
| AX T1 SE | 12 | 450 | 2:26 | 150 | 3.04 | 1.40E-02 |
| AX T2 TSE | 94 | 3800 | 2:22 | 150 | 3.71 | 7.20E-02 |
| | | | | | | 3.39E-02 |
| Chest Landmark | | | | | | |
| Protocol | TE (ms) | TR (ms) | TA (min) | FA (°) | B1+ (µT) | $CEM_{43}$ |
| COR TRUFI | 1.21 | 491.33 | 0:20 | 64 | 4.01 | 3.92E-01 |
| AX TRUFI | 1.18 | 328.38 | 0:13 | 64 | 3.99 | 7.19E-02 |
| COR VIBE | 2.39, 4.77 | 6.63 | 0:22 | 10 | 1.28 | 2.08E-04 |
| AX VIBE | 2.39, 4.77 | 6.9 | 0:20 | 10 | 1.49 | 2.44E-04 |
| TRUFI CINE 2C | 1.16 | 38.22 | 0:06 | 60 | 3.98 | 2.20E-03 |
| PC 3C IP AORTA | 2.47 | 37.12 | 0:10 | 20 | 0.97 | 5.27E-05 |
| PC RVOT IP | 2.47 | 37.12 | 0:10 | 20 | 0.97 | 5.27E-05 |
| PC TP AV | 2.47 | 37.12 | 0:10 | 20 | 0.97 | 5.27E-05 |
| TRUFI CINE SA | 1.16 | 40.96 | 1:00 | 60 | 3.99 | 3.71E+01 |
| TFL PSIR | 3.22 | 700 | 0:14 | 25 | 0.63 | 6.65E-05 |
| Abdomen Landmark | | | | | | |
| Protocol | TE (ms) | TR (ms) | TA (min) | FA (°) | B1+ (µT) | $CEM_{43}$ |
| T2 HASTE COR MBH | 91 | 1400 | 0:48 | 180 | 4.2 | 1.73E-01 |
| T2 HASTE FS TRA MBH | 94 | 1400 | 0:05 | 160 | 3.1 | 4.70E-05 |
| T2 BLADE FS TRA | 91 | 2200 | 2:35 | 160 | 3.35 | 8.14E-01 |
| T2 TSE FS TRA MBH | 86 | 4780 | 0:54 | 160 | 3.77 | 6.57E-02 |
| EP2D DIFF | 54 | 6200 | 3:18 | - | 1.3 | 2.33E-03 |

Table 3 continued. Sequence Parameters and scanner-reported RMS $B_1^+$ values for different imaging protocols at 3 T scanner.

| Protocol | TE (ms) | TR (ms) | TA (min) | FA (°) | B1+ (µT) | CEM$_{43}$ |
|---|---|---|---|---|---|---|
| **3T Imaging** | | | | | | |
| **Head Landmark** | | | | | | |
| T1-TSE DARK-FLUID | 8.5 | 2000 | 4:38 | 150 | 1.34 | 6.21E-03 |
| T1-SPACE | 9 | 700 | 6:41 | - | 1.81 | 5.32E-02 |
| T1-FL2D-TRA | 2.49 | 250 | 1:19 | 70 | 1.97 | 4.47E-03 |
| T1-FL2D_COR | 2.49 | 250 | 1:19 | 70 | 2.25 | 1.04E-02 |
| T2-TSE-TRA | 100 | 6000 | 2:44 | 150 | 1.99 | 1.90E-02 |
| T2-TSE-DARK-FLUID | 83 | 9000 | 4:14 | 150 | 1.46 | 7.52E-03 |
| EP2D-DIFF | 81 | 3700 | 0:57 | - | 1.51 | 8.98E-04 |
| TOF-FL3D | 3.42 | 21 | 5:33 | 18 | 2.02 | 8.81E-02 |
| T1-SPACE-FS | 10 | 750 | 7:00 | - | 2.15 | 2.65E-01 |
| **Chest Landmark** | | | | | | |
| T1-TSE-DB | 27 | 700 | 0:09 | 180 | 1.55 | 1.17E-04 |
| T2-TSE-DB | 71 | 800 | 0:05 | 180 | 1.85 | 6.42E-05 |
| T2-TSE-DB-FAST-HR | 70 | 600 | 0:04 | 180 | 2.2 | 6.78E-05 |
| T2-TRIM-DB-SAX | 44 | 800 | 0:06 | 180 | 1.98 | 1.10E-04 |
| T2-TSE-BLADE-DB | 76 | 750 | 2:05 | 180 | 1.19 | 5.54E-03 |
| HASTE-16-SL-DB-PACE | 49 | 750 | 1:20 | 160 | 1.39 | 5.46E-03 |
| TRUFI-SINGLESHOT | 1.23 | 224.2 | 0:12 | 60 | 1.95 | 4.73E-04 |
| DE-OVERVIEW-TFI-PSIR | 1.09 | 700 | 0:18 | 55 | 1.6 | 4.78E-04 |
| **Abdomen Landmark** | | | | | | |
| T2-HASTE-COR-MBH | 87 | 1400 | 0:48 | 160 | 2.27 | 3.91E-01 |
| T2-TSE-FS-TRA | 100 | 2200 | 3:25 | 160 | 2.04 | 3.30E+00 |
| T2-BLADE-FS-TRA | 89 | 2500 | 3:30 | 135 | 2.48 | 1.10E+01 |
| EP2D-DIFF-TRA | 39 | 4600 | 3:28 | - | 1.89 | 1.47E+00 |
| T1-VIBE-DIXON-TRA | 1.29, 2.52 | 3.97 | 0:15 | 9 | 1.52 | 2.64E-04 |
| T2-SPACE-COR | 701 | 2400 | 4:46 | 120 | 2.1 | 1.01E+01 |

Table 4. Gradient-induced extrinsic potential

| FRL # | Length [cm] | $V_{emf}$ [V] | $E_{tissue}$ (V/m) |
|---|---|---|---|
| 1 | 24.0 | 2.84 | 23.68 |
| 2 | 14.2 | 1.97 | 16.44 |
| 3 | 13.4 | 1.90 | 12.68 |
| 4 | 18.0 | 2.31 | 76.99 |
| 5 | 17.2 | 2.24 | 17.22 |
| 6 | 15.5 | 2.09 | 13.92 |
| 7 | 2.0 | 0.32 | 64.00 |
| 8 | 4.5 | 0.72 | 16.74 |
| 9 | 24.6 | 2.90 | 24.13 |
| 10 | 8.5 | 1.36 | 17.44 |

**Supplemental Materials**

Supporting Information Table S1. Temperature rise $\Delta T$ [°C] in the tissue surrounding the FRL after 10-minute continuous RF exposure at 64 MHz (1.5 T) for the coil iso-center positioned at different imaging landmarks and the input power adjusted to generate different B1+ values on an axial plane passing through center of the coil.

| FRL # | $B_1^+$ [$\mu T$] | | | | | Landmark |
|---|---|---|---|---|---|---|
| | 1 | 2 | 3 | 4 | 5 | |
| 1 | 0.14 | 0.57 | 1.28 | 2.27 | 3.55 | Abdomen |
| | 0.23 | 0.94 | 2.11 | 3.75 | 5.87 | Chest |
| | 0.09 | 0.37 | 0.84 | 1.49 | 2.32 | Head |
| 2 | 0.05 | 0.22 | 0.49 | 0.87 | 1.36 | Abdomen |
| | 0.18 | 0.74 | 1.66 | 2.95 | 4.61 | Chest |
| | 0.10 | 0.42 | 0.95 | 1.69 | 2.64 | Head |
| 3 | 0.12 | 0.49 | 1.10 | 1.96 | 3.06 | Abdomen |
| | 0.23 | 0.70 | 2.11 | 3.76 | 5.87 | Chest |
| | 0.11 | 0.43 | 0.96 | 1.70 | 2.66 | Head |
| 4 | 0.10 | 0.41 | 0.91 | 1.62 | 2.53 | Abdomen |
| | 0.26 | 1.02 | 2.30 | 4.09 | 6.38 | Chest |
| | 0.11 | 0.42 | 0.96 | 1.70 | 2.66 | Head |
| 5 | 0.08 | 0.31 | 0.70 | 1.24 | 1.94 | Abdomen |
| | 0.29 | 1.16 | 2.60 | 4.62 | 7.22 | Chest |
| | 0.14 | 0.55 | 1.23 | 2.18 | 3.40 | Head |
| 6 | 0.94 | 3.76 | 8.46 | 15.04 | 23.50 | Abdomen |
| | 2.01 | 8.03 | 18.06 | 32.12 | 50.18 | Chest |
| | 0.53 | 2.14 | 4.81 | 8.55 | 13.35 | Head |
| 7 | 0.16 | 0.65 | 1.45 | 2.58 | 4.03 | Abdomen |
| | 0.40 | 1.59 | 3.57 | 6.35 | 9.93 | Chest |
| | 0.03 | 0.44 | 0.98 | 1.74 | 2.72 | Head |
| 8 | 0.06 | 0.23 | 0.52 | 0.93 | 1.46 | Abdomen |
| | 0.05 | 0.20 | 0.45 | 0.81 | 1.26 | Chest |
| | 0.06 | 0.26 | 0.57 | 1.02 | 1.60 | Head |
| 9 | 0.10 | 0.39 | 0.89 | 1.58 | 2.46 | Abdomen |
| | 0.43 | 1.73 | 3.90 | 6.94 | 10.84 | Chest |
| | 0.14 | 0.56 | 1.25 | 2.23 | 3.48 | Head |
| 10 | 0.05 | 0.22 | 0.48 | 0.86 | 1.34 | Abdomen |
| | 0.04 | 0.14 | 0.32 | 0.57 | 0.90 | Chest |
| | 0.14 | 0.57 | 1.27 | 2.26 | 3.54 | Head |

Supporting Information Table S2. Temperature rise $\Delta T$ [°C] in the tissue surrounding the FRL after 10-minute continuous RF exposure at 127 MHz (3 T) for the coil iso-center positioned at different imaging landmarks and the input power adjusted to generate different $B_1^+$ values on an axial plane passing through center of the coil.

| FRL # | $B_1^+$ [µT] | | | | | Landmark |
|---|---|---|---|---|---|---|
| | 1 | 2 | 3 | 4 | 5 | |
| 1 | 0.26 | 1.02 | 2.30 | 4.09 | 6.39 | Abdomen |
| | 1.25 | 4.98 | 11.20 | 19.91 | 31.11 | Chest |
| | 1.16 | 4.64 | 10.44 | 18.55 | 28.99 | Head |
| 2 | 0.26 | 1.03 | 2.32 | 4.13 | 6.46 | Abdomen |
| | 0.42 | 1.67 | 3.76 | 6.69 | 10.45 | Chest |
| | 0.37 | 1.49 | 3.35 | 5.95 | 9.30 | Head |
| 3 | 1.94 | 7.74 | 17.42 | 30.97 | 48.40 | Abdomen |
| | 2.46 | 9.82 | 22.10 | 39.29 | 61.39 | Chest |
| | 0.78 | 3.13 | 7.04 | 12.51 | 12.97 | Head |
| 4 | 2.24 | 8.94 | 20.12 | 35.77 | 55.90 | Abdomen |
| | 0.28 | 1.12 | 2.52 | 4.48 | 6.99 | Chest |
| | 0.26 | 1.03 | 2.31 | 4.10 | 6.41 | Head |
| 5 | 0.28 | 2.07 | 2.55 | 4.53 | 7.08 | Abdomen |
| | 1.43 | 5.70 | 12.83 | 22.81 | 35.65 | Chest |
| | 0.53 | 2.11 | 4.75 | 8.45 | 13.21 | Head |
| 6 | 0.37 | 1.47 | 3.32 | 5.90 | 9.21 | Abdomen |
| | 2.07 | 8.27 | 18.62 | 33.09 | 51.71 | Chest |
| | 0.73 | 2.94 | 6.61 | 11.75 | 18.37 | Head |
| 7 | 0.27 | 1.07 | 2.30 | 4.28 | 6.68 | Abdomen |
| | 1.18 | 4.72 | 10.62 | 18.89 | 29.51 | Chest |
| | 0.13 | 0.51 | 1.15 | 2.05 | 3.20 | Head |
| 8 | 0.22 | 0.88 | 1.98 | 3.51 | 5.49 | Abdomen |
| | 0.07 | 0.28 | 0.63 | 1.12 | 1.74 | Chest |
| | 0.13 | 0.53 | 1.20 | 2.13 | 3.32 | Head |
| 9 | 0.81 | 3.24 | 7.29 | 12.96 | 20.25 | Abdomen |
| | 1.53 | 6.11 | 13.75 | 24.44 | 38.19 | Chest |
| | 0.99 | 3.98 | 8.96 | 15.93 | 24.88 | Head |
| 10 | 0.27 | 1.09 | 2.44 | 4.34 | 6.78 | Abdomen |
| | 0.09 | 0.37 | 0.83 | 1.47 | 2.30 | Chest |
| | 0.10 | 0.40 | 0.90 | 1.60 | 2.49 | Head |

Supporting Information Table S3. Lead length factor L.

| Lead length [cm] | $V_{emf}$ [Volt] |
|---|---|
| $l \leq 10$ | $16 \times 10^{-2} \times l$ |
| $10 < l < 63$ | $8.87 \times 10^{-2} \times l - 7.13 \times 10^{-1}$ |

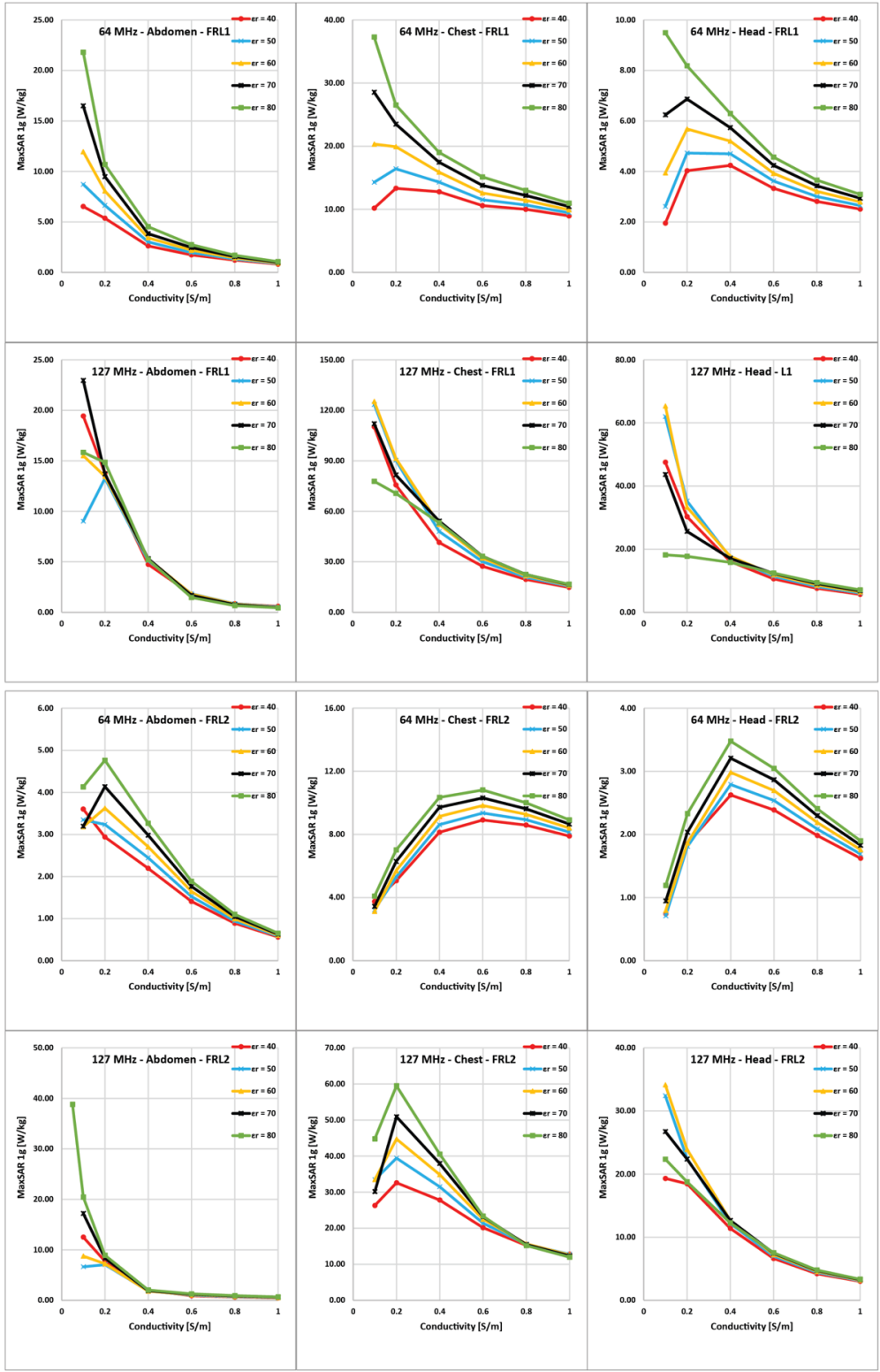

Supporting Information Figure S1. MaxSAR1g generated around retained lead models FRL1 and FRL2 as a function of body model's conductivity (horizontal axis) and permittivity (different-colored graphs). The field strength and imaging landmark is noted on top of each plot.

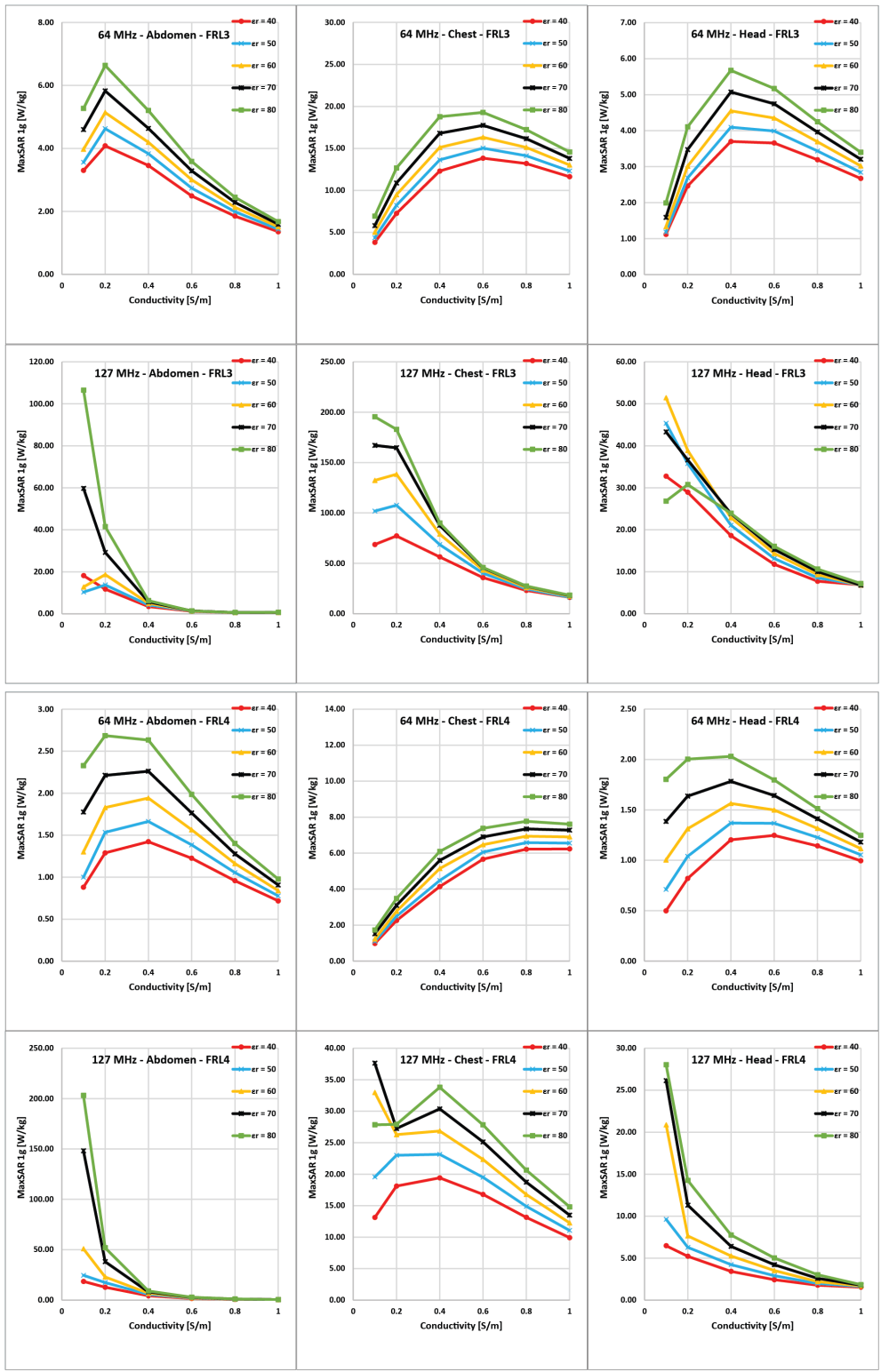

Supporting Information Figure S2. MaxSAR1g generated around retained lead models FRL3 and FRL4 as a function of body model's conductivity (horizontal axis) and permittivity (different-colored graphs). The field strength and imaging landmark is noted on top of each plot.

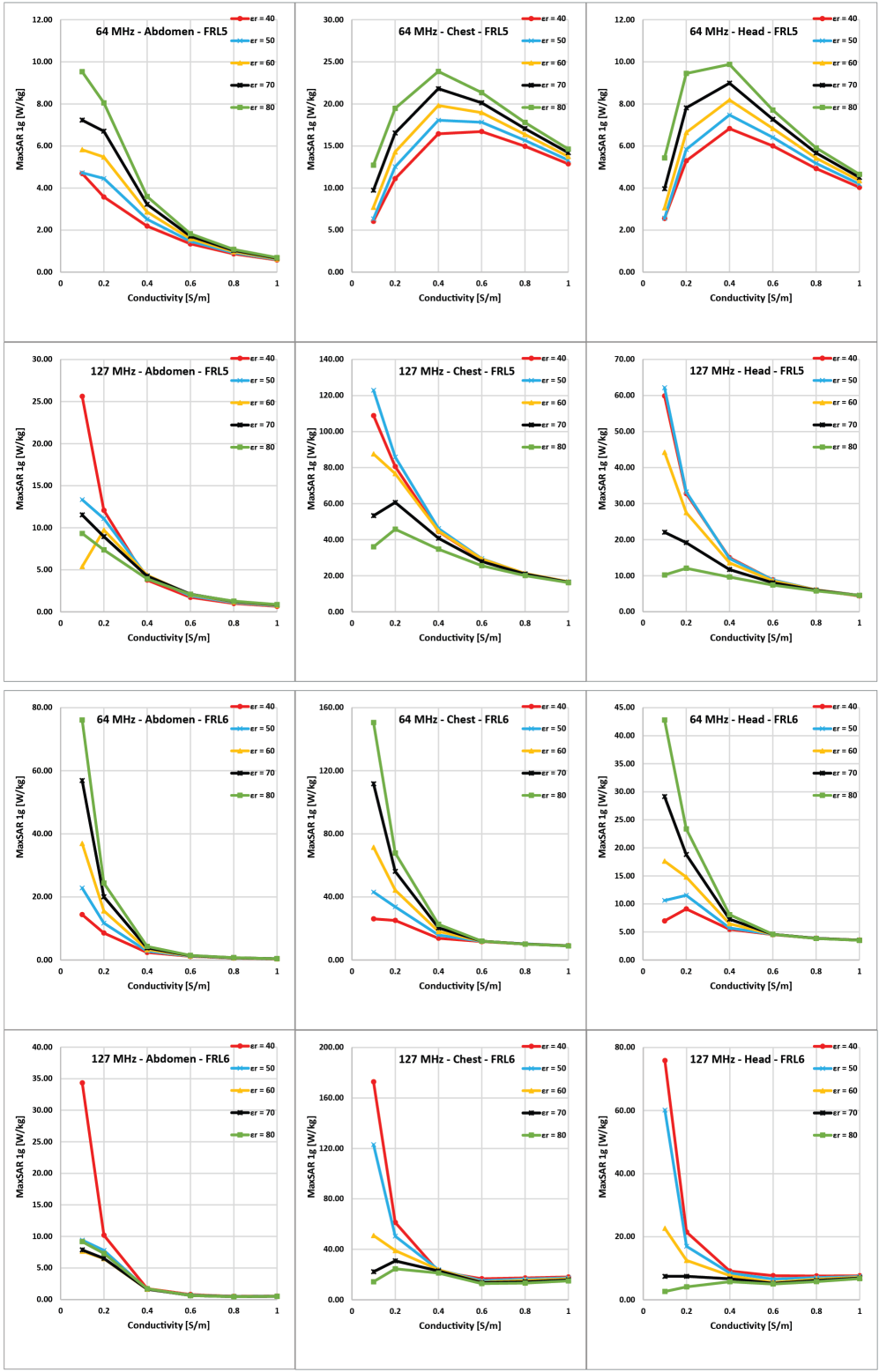

Supporting Information Figure S3. MaxSAR1g generated around retained lead models FRL5 and FRL6 as a function of body model's conductivity (horizontal axis) and permittivity (different-colored graphs). The field strength and imaging landmark is noted on top of each plot.

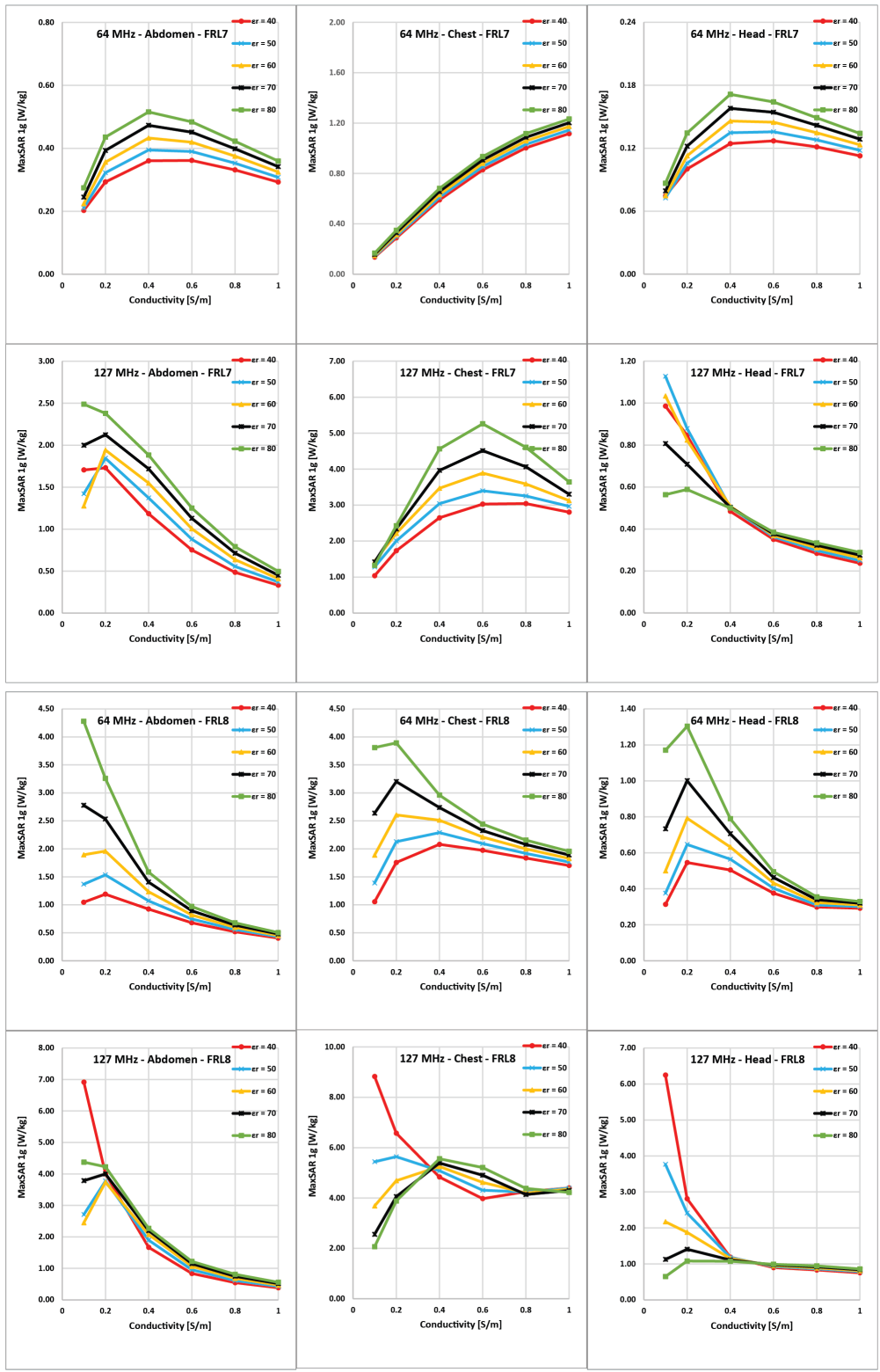

Supporting Information Figure S4. MaxSAR1g generated around retained lead models FRL7 and FRL8 as a function of body model's conductivity (horizontal axis) and permittivity (different-colored graphs). The field strength and imaging landmark is noted on top of each plot.

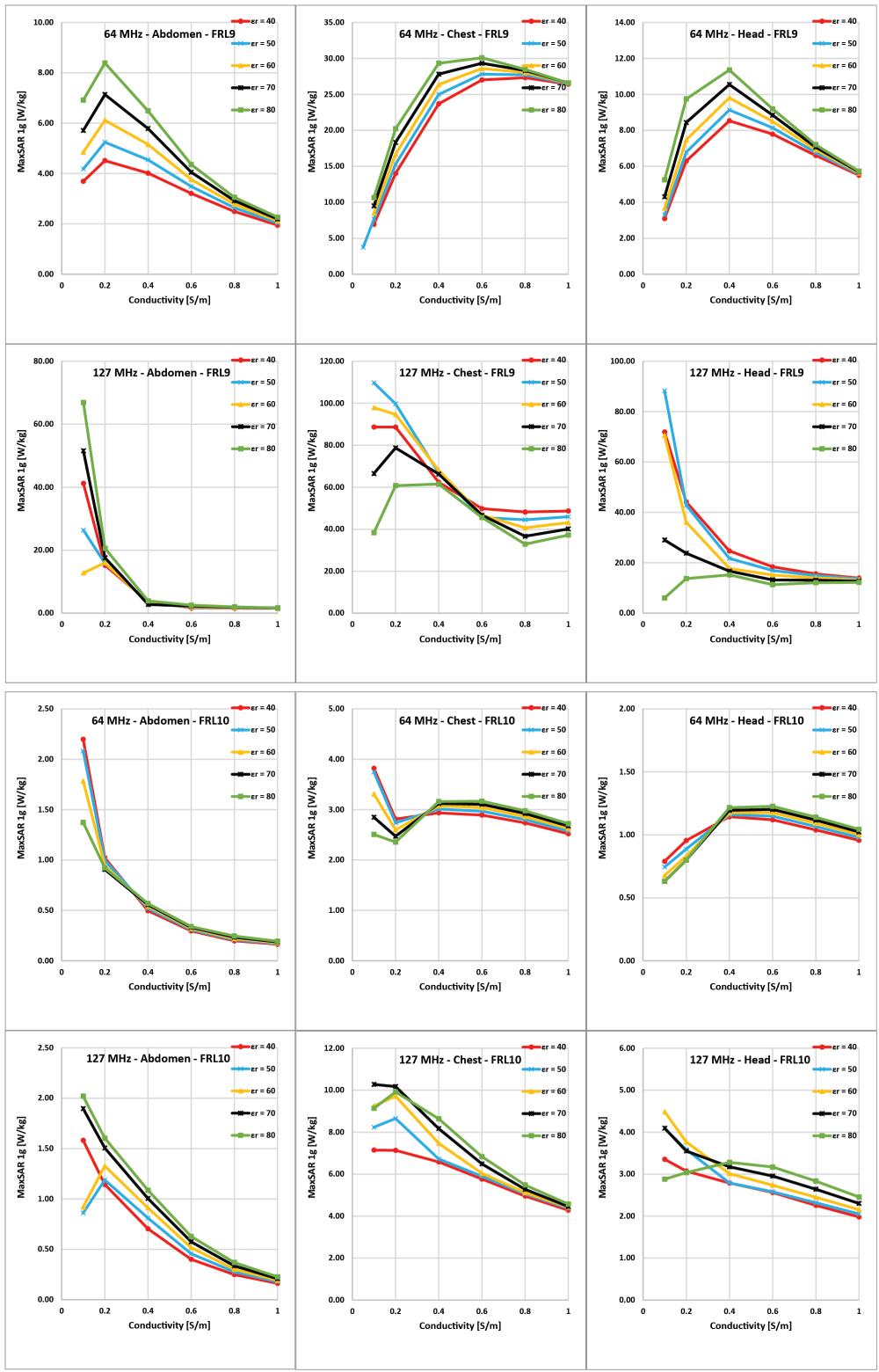

Supporting Information Figure S5. MaxSAR1g generated around retained lead models FRL9 and FRL10 as a function of body model's conductivity (horizontal axis) and permittivity (different-colored graphs). The field strength and imaging landmark is noted on top of each plot.

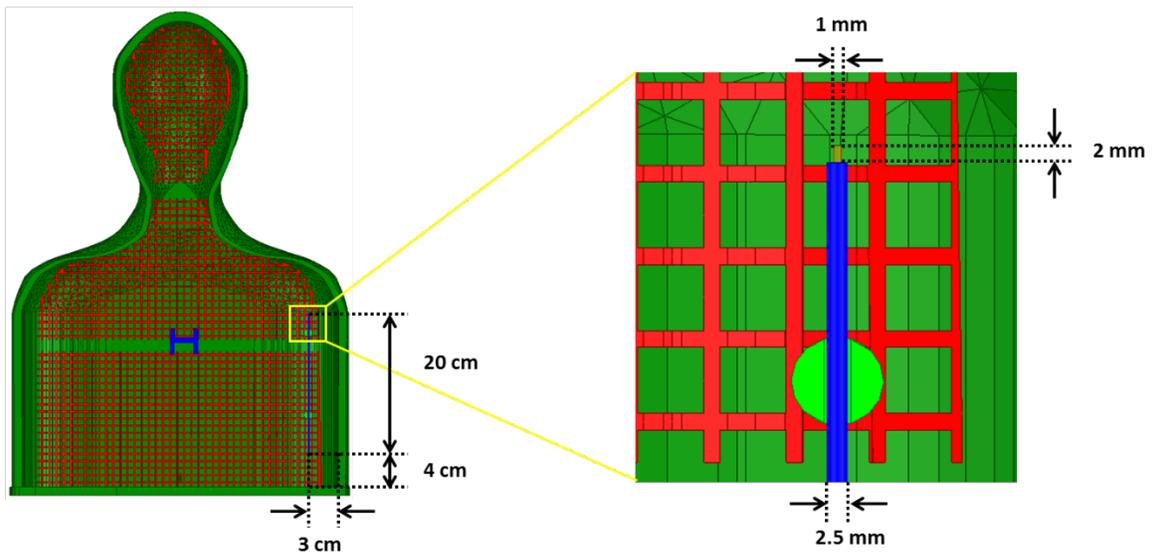
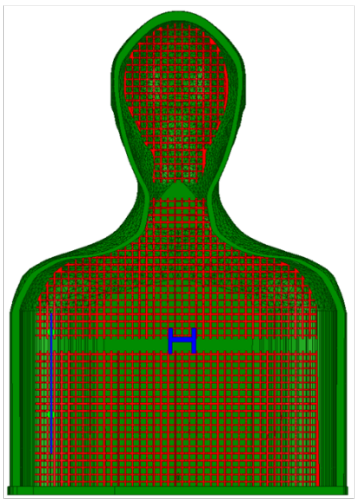
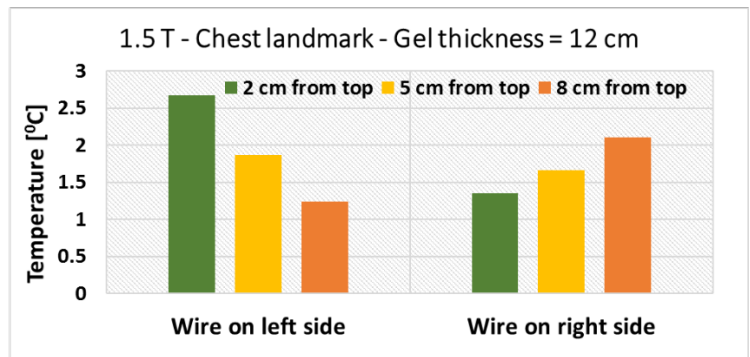

Supporting Information Figure S6. Measured temperature rise along length of 20 cm wire at 1.5 T. The wire was located at the left and the right of the phantom, and the depth of the wire was 2 cm, 5 cm and 8 cm from the top of the gel (maximum thickness = 12 cm).

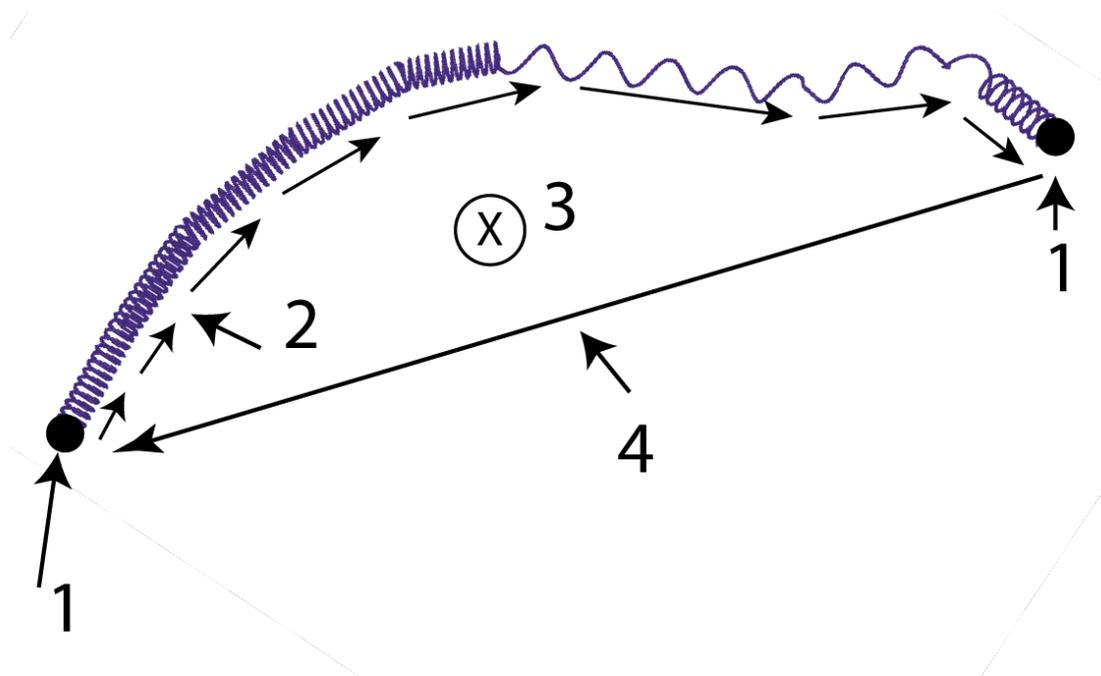

Supporting Information Figure S7. (1): Two ends of the FRL. (2) Tangential component of gradient-induced electric field along the FRL. A conservative estimation of the induced voltage $V_{emf}$ along the FRL was calculated by multiplying the maximum value of gradient-induced E field (based on simulations given in Annex B of ISO-TS 10974) by the FRL length. (3) Gradient field (4) A conservative estimation of E field in the tissue was calculated by dividing $V_{emf}$ by the distance between two ends of the lead.